\documentstyle[12pt]{article}
\input{psfig}
\long\def\comment#1{}
\begin{document}
\title{The Speed of Light and Pur\={a}\d{n}ic Cosmology}

\author{Subhash Kak\\
Department of Electrical \& Computer Engineering\\
Louisiana State University\\
Baton Rouge, LA 70803-5901, USA\\
FAX: 225.578.5200; Email: {\tt kak@ee.lsu.edu}}

\date{April 18, 1998 {\small Corrected Dec 30, 2000}}
\maketitle

\begin{abstract}
We survey early Indian ideas on the speed of light
and the size of the universe.
A context is provided for S\={a}ya\d{n}a's
statement (14th century) that the speed is 2,202 yojanas per
half nime\d{s}a (186,000 miles per second!). 
It is shown how this statement may have emerged from
early
Pur\={a}\d{n}ic notions
regarding the size of the universe.
Although this value can only be considered to be an amazing
coincidence, the Pur\={a}\d{n}ic cosmology at the basis of
this assertion illuminates many ancient ideas of space and
time.

{\it Keywords:} Speed of light, Ancient Indian astronomy, Pur\={a}\d{n}ic cosmology
\end{abstract}


Indian texts consider 
light to be like a wind. 
Was any thought 
given to its speed?
Given the nature of the analogy, one would 
expect that this speed was considered finite.
The Pur\={a}\d{n}as speak of the moving {\it
jyoti\'{s}cakra}, ``the circle of light.''
This analogy or that of the swift arrow let loose from the
bow in these accounts leaves ambiguous whether
the circle of light is the Sun or its speeding rays.

We get a specific number 
that could refer to the speed of light
in a medieval text by
S\={a}ya\d{n}a (c. 1315-1387), prime minister in the court of
Emperors Bukka I and his successors of the Vijayanagar Empire and Vedic scholar.
In his commentary on the fourth verse
of the hymn 1.50 of the \d{R}gveda on the Sun, he says$^1$
 
\begin{quote}
{\it tath\={a} ca smaryate yojan\={a}n\={a}\d{m} sahasre dve dve \'{s}ate dve ca 
yojane
ekena nimi\d{s}\={a}rdhena kramam\={a}\d{n}a}\\
 
Thus it is remembered:  [O Sun] you who
traverse 2,202 {\it yojanas} in half a {\it nime\d{s}a}.
\end{quote}

The same statement occurs in the commentary on the
Taittir\={\i}ya Br\={a}hma\d{n}a by Bha\d{t}\d{t}a
Bh\={a}skara (10th century?), where it is said to
be an old Pur\={a}\d{n}ic tradition.
 
The figure could refer to the actual motion of the Sun
but, as we will see shortly, that is impossible.
Is it an old tradition related to the speed of [sun]light that
S\={a}ya\d{n}a appears to suggest?
We would like to know if that supposition is true by
examining parallels in the Pur\={a}\d{n}ic literature.

The units of {\it yojana} and {\it nime\d{s}a}
are well known. The usual meaning of yojana is about
9 miles as in
the {\it Artha\'{s}\={a}stra} where it is
defined as being equal to 8,000 {\it dhanu} or ``bow,'' where
each dhanu is taken to be about
6 feet. \={A}ryabha\d{t}a, Brahmagupta and other astronomers used
smaller yojanas but such exceptional usage was confined to
the astronomers;
we will show that
the Pur\={a}\d{n}as also use a non-standard measure of
yojana.
As a scholar of the Vedas 
and a non-astronomer, 
S\={a}ya\d{n}a would be expected to use the ``standard''
Artha\'{s}\={a}stra units.

The measures of time are thus defined in the Pur\={a}\d{n}as:

\begin{quote}

15 nime\d{s}a = 1 k\={a}\d{s}\d{t}h\={a}

30 k\={a}\d{s}\d{t}h\={a} = 1 kal\={a}

30 kal\={a} = 1 muh\={u}rta

30 muh\={u}rta = 1 day-and-night

\end{quote}

A nime\d{s}a is therefore equal to $\frac{16}{75}$ seconds.

De and Vartak have in recent books$^2$ argued that
this statement refers to the speed of light.
Converted into modern units, it does come very close
to the correct figure of 186,000 miles per second!

Such an early knowledge of this number
doesn't sound credible because the speed
of light was determined only in 1675 by Roemer who looked at
the difference in the times that light from Io, one of the moons of
Jupiter, takes to reach Earth based on whether it is on the near side
of Jupiter or the far side.  Until then light was taken to travel with
infinite velocity. 
There is no record of any optical experiments
that could have been performed in India before the modern 
period to measure the speed of light.
 
Maybe S\={a}ya\d{n}a's figure
refers to the speed of the Sun in its supposed orbit around the Earth.
But that places the orbit of the Sun at a distance of over 2,550
million miles.  The correct value is only 93 million miles and until
the time of Roemer the distance to the Sun used to be taken to be less
than 4 million miles.  This interpretation takes us nowhere.
The Indian astronomical texts place the Sun only about half a million
yojanas from the Earth.

What about the possibility of fraud?  S\={a}ya\d{n}a's statement was printed in
1890 in the famous edition of \d{R}gveda edited by Max M\"{u}ller, the German
Sanskritist.  He claimed to have used several three or four hundred
year old manuscripts of S\={a}ya\d{n}a's commentary, written much before the
time of Roemer.
 
Is it possible that M\"{u}ller was duped by an Indian correspondent who
slipped in the line about the speed?  Unlikely, because S\={a}ya\d{n}a's
commentary is so well known that an interpolation would have been long
discovered.  And soon after M\"{u}ller's ``Rigveda'' was published, someone
would have claimed that it contained this particular ``secret''
knowledge. Besides,
a copy of S\={a}ya\d{n}a's commentary, dated 1395,
is preserved in the Central Library, Vadodara.$^3$

One can dismiss S\={a}ya\d{n}a's number as a meaningless coincidence.
But that would be a mistake if there exists a framework of 
ideas---an old physics---in which this
number makes sense.
We explore the prehistory of this number by
considering early textual
references.
We will show that these references in the Pur\={a}\d{n}as
and other texts indicate that S\={a}ya\d{n}a's speed
is connected, numerically, to very ancient ideas.
This helps us understand the framework of ideas
regarding the
universe that led to this figure.

\section*{Physical ideas in the Indian literature}

The Vedas take the universe to be infinite in size. 
The universe was visualized in the image of the
cosmic egg, {\it Brahm\={a}\d{n}\d{d}a}.
Beyond our own universe lie other universes. 

The Pa\~{n}cavi\d{m}\'{s}a
Br\={a}hma\d{n}a 16.8.6 states that the heavens are
1000 earth diameters away
from the Earth.
The Sun was taken to be halfway to the heavens, so
this suggests a distance to the Sun to be about
500 earth diameters from the Earth, which is
about 0.4375 million yojanas.

Yajurveda, in the mystic hymn 17, dealing with the nature of the universe,
counts numbers in powers of
ten upto $10^{12}$. It has been suggested that this 
is an estimate of the size of the universe in yojanas.

The philosophical schools of 
S\={a}\d{m}khya and
Vai\'{s}e\d{s}ika tell us about the old ideas on
light.$^4$
According to 
S\={a}\d{m}khya, light is one of the five fundamental
``subtle'' elements ({\it tanm\={a}tra}) out of which
emerge the gross elements.
The atomicity of 
these elements is not specifically mentioned and it appears
that they were actually taken to be continuous.

On the other hand, 
Vai\'{s}e\d{s}ika
is an atomic theory of the physical
world on the nonatomic ground of ether, space and time.
The basic atoms are those of 
earth ({\it p\d{r}thiv\={\i}}), water ({\it \={a}pas}), fire
({\it tejas}), and air ({\it v\={a}yu}), that should
not be confused with the ordinary meaning of these terms.
These atoms are taken to form binary molecules that 
combine further to form larger molecules.$^5$
Motion is defined in terms of the movement of the
physical atoms and it appears that it is taken to be
non-instantaneous.  

Light rays are taken to be a stream of
high velocity of tejas atoms.
The particles of light
can exhibit different characteristics
depending on the speed and the
arrangements of the tejas atoms.
 
Although there 
existed several traditions of astronomy in India,$^6$ only the
mathematical astronomy of the Siddh\={a}ntas has been
properly examined. 
Some of the information of the non-Siddh\={a}ntic
astronomical systems is preserved in the
Pur\={a}\d{n}as.

The Pur\={a}\d{n}ic astronomy is cryptic, and since
the Pur\={a}\d{n}as are encyclopaedic texts, with
several layers of writing, presumably by different
authors, there are inconsistencies
in the material.
Sometimes, speculative and the empirical ideas are so
intertwined that without care the material can
appear meaningless.
The Pur\={a}\d{n}ic geography is quite fanciful and this
finds parallels in its astronomy as well.

We can begin the process of understanding
Pur\={a}\d{n}ic astronomy by considering its
main features, such as the size of the solar system
and the motion of the Sun. 
But before we do so, we will speak briefly of
the notions in the Siddh\={a}ntas.


\section*{Size of the universe in the
\={A}ryabha\d{t}\={\i}ya}

\={A}ryabha\d{t}a 
in his {\it \={A}ryabha\d{t}\={\i}ya (AA)}
deals with the question of the size of the universe.
He defines a
{\it yojana} to be
8,000 {\it n\d{r}},
where a {\it n\d{r}} is the height of a man;
this makes his yojana ($y_a$) approximately 7.5 miles.$^{7}$
Or $y_s \approx \frac{6}{5} y_a$,
where $y_s$ is the standard Artha\'{s}\={a}stra
yojana.
AA 1.6
states that the orbit of the Sun is
2,887,666.8 {\it yojanas} and that of the
sky is 12,474,720,576,000 {\it yojanas}.

Commenting on this, Bh\={a}skara I (c. 629) says:

\begin{quote}

{\it y\={a}vantam\={a}k\={a}\'{s}aprade\'{s}am ravermay\={u}kh\={a}\d{h}
samant\={a}t dyotayanti t\={a}v\={a}n prade\'{s}a\d{h}
khagolasya paridhi\d{h} khakak\d{s}y\={a}.
anyath\={a} hyaparimitatv\={a}t \={a}k\={a}\'{s}asya
parim\={a}\d{n}\={a}khy\={a}nam nopapadyate.}\\

That much of the sky as the Sun's rays illumine on all sides is
called the orbit of the sky. Otherwise, the sky is beyond limit;
it is impossible to state its measure.$^8$

\end{quote}

This implies that while the universe is infinite, the
solar system extends as far as the rays of the Sun can
reach.

There is no mention by \={A}ryabha\d{t}a of a
speed of light. But 
the range of light particles is taken to be finite,
so it must have been assumed that the particles in the
``observational universe'' do not penetrate to
the regions beyond the ``orbit of the sky.''
This must have been seen in the 
analogy of the gravitational pull
of the matter just as other particles fall back
on Earth after reaching a certain height.

The orbit of the sky is $4.32 \times 10^6$ greater than the
orbit of the Sun.
It is clear
that this enlargement was inspired by cosmological ideas.

The diameters of the Earth, the Sun, and the Moon are taken to
be 1,050, 4,410 and 315 yojanas, respectively.
Furthermore, {\it AA} 1.6 implies
the distance to the Sun, $R_s$, 
to be 459,585 yojanas, and that to the Moon, $R_m$, as 34,377 yojanas.
These distances are in the correct proportion related to
their assumed sizes given that the distances are
approximately 108 times the corresponding diameters.$^9$

Converted to the standard {\it Artha\'{s}\={a}stra} units, the diameters
of the Earth and the Sun are about 875 and 3,675 yojanas,
and the distance to the Sun
is around 0.383 million yojanas.

\={A}ryabha\d{t}a considers the 
orbits, with respect to the Earth, in the
correct order Moon, Mercury, Venus, Sun, Mars, Jupiter, and Saturn,
based on their periods.

\section*{Pur\={a}\d{n}ic cosmology}

The Pur\={a}\d{n}as provide material which is
believed to be closer to the knowledge
of the Vedic times.$^{10}$
Here we specifically consider 
V\={a}yu Pur\={a}\d{n}a (VaP), 
Vi\d{s}\d{n}u Pur\={a}\d{n}a (ViP),
and Matsya Pur\={a}\d{n}a (MP).
VaP and ViP are generally believed to be amongst the earliest
Pur\={a}\d{n}as and at least 1,500 years old.
Their astronomy is prior to the Siddh\={a}ntic
astronomy of \={A}ryabha\d{t}a and his successors.

The Pur\={a}\d{n}as instruct through myth and 
this mythmaking can be seen in their approach to
astronomy also.
For example, they speak of seven underground
worlds
below the orbital plane of the planets
and of seven ``continents'' encircling the Earth.
One has to take care to separate this
imagery, that parallels the conception
of the seven centres of the human's
psycho-somatic body, from the underlying cosmology 
of the Pur\={a}\d{n}as
that is their primary concern in their {\it jyoti\d{s}a}
chapters.

It should be noted that the idea of seven regions of
the universe is present in the \d{R}gveda 1.22.16-21 where
the Sun's stride is described as {\it saptadh\={a}man},
or taking place in seven regions.

The different Pur\={a}\d{n}as appear to 
reproduce the same cosmological material.
There are some minor differences in 
figures that may be a result of wrong 
copying by scribes who did not understand the
material.
In this paper, we mainly follow ViP.

ViP 2.8 describes the Sun to be 9,000 yojanas in
length and to be connected by an axle that is
$15.7 \times 10^6$ yojanas long to the 
M\={a}nasa mountain and another axle
45,500 yojanas long connected to the pole star.
The distance of 15.7 million yojanas between the
Earth and the Sun is much greater than the
distance of 0.38 or 0.4375 million yojanas that we
find in the Siddh\={a}ntas and other early books.
This greater distance is stated without a corresponding change in
the diameter of the Sun.
It is interesting that this distance is less than one
and a half times the correct value; the value of
the Siddh\={a}ntas is one-thirtieth the correct value.

Elsewhere, in VaP 50, it is stated that the Sun
covers 3.15 million yojanas in a muh\={u}rta.
This means that the distance covered in a day are 94.5
million yojanas.
MP 124 gives the same figure.
This is in agreement with the view that the Sun is
15.7 million yojanas away from the Earth.
The specific speed given here, translates to 
116.67 yojanas per half-nime\d{s}a.

The size of the universe is described in two different
ways, through the ``island-continents'' and through
heavenly bodies.

The geography of the Pur\={a}\d{n}as describes a central
continent, Jambu, surrounded by alternating bands of ocean and land.
The seven island-continents 
of Jambu, Plak\d{s}a, \'{S}\={a}lmala, Ku\'{s}a,
Kraunca, \'{S}\={a}ka, and Pu\d{s}kara are encompassed, successively,
by seven oceans; and each ocean and continent is, respectively,
of twice the extent of that which precedes it.
The universe is seen as a sphere of size 500 million yojanas.

It is important to realize that the continents are
imaginary regions and they should not be
confused with the continents on the Earth.
Only certain part of the innermost planet, Jambu, that deal with
India have
parallels with
real geography.

The inner continent is taken to be 
16,000 yojanas as the base of the world axis.
In opposition to the interpretation by
earlier commentators, who took the
increase in dimension by a factor of
two is only across the seven ``continents,''
we take it to apply to the ``oceans'' as well. 
We have done this because it harmonizes many numbers and
so it appears to have been a plausible model that led
to the development of the system.
In itself, it has no bearing on the question of the
speed of light that we will discuss later.

At the end of the seven island-continents is a region that is twice the preceding
region. Further on, is the Lok\={a}loka mountain, 10,000 yojanas in breadth,
that marks the end of our universe.

Assume that the size of the Jambu is $J$ yojana,
then the size of the universe is:

\begin{equation}
U = J ( 1 +2 +2^2+ 2^3 +2^4 +2^5 +2^6 +2^7 +2^8 +2^9 +2^{10} +2^{11} +2^{12} +2^{13} +2^{14}) +20,000 
\end{equation}

Or,

\begin{equation}
U =  32,767 J + 20,000~ yojanas
\end{equation}

If U is 500 million yojanas, then J should be about 15,260 yojanas.
The round figure of 16,000 is mentioned as the width of the base of
the Meru, the world axis, at the surface of the Earth.
This appears to support our interpretation.
This calculation assumes that around the Meru of size 16,000
yojanas is the rest of the Jambu continent which circles
another 16,000 yojanas. In other words, it takes the
diameter of Jambu to be about 48,000 yojanas.

Note that the whole description of the Pur\={a}\d{n}ic cosmology
had been thought to be inconsistent because 
an erroneous interpretation of the increase in the sizes of
the ``continents'' had been used.

When considered in juxtaposition with the preceding numbers,
the geography of concentric continents is
a representation of the plane of the Earth's rotation, with
each new continent as the orbit of the next ``planet''.$^{11}$

The planetary model in the Pur\={a}\d{n}as is
different from that in the Siddh\={a}ntas.
Here the Moon as well as the planets are in orbits
higher than the Sun. 
Originally, this supposition for the Moon may
have represented the fact that it goes higher than the
Sun in its orbit.
Given that the Moon's inclination is $5^\circ$ to
the ecliptic, its declination can be $28.5^\circ$
compared to the Sun's maximum declination of
$\pm 23.5^\circ$.
This ``higher'' position must have been, at
some stage, represented literally by a
higher orbit. To make sense with the
observational reality, it became necessary for
the Moon is taken to be twice as large as
the Sun. 

The distances of the planetary orbits beyond the Sun are
as follows:

\newpage
Table 1: From Earth to Pole-star

\begin{tabular}{||l|r||} \hline
Interval I & yojanas\\ \hline
Earth to Sun & 15,700,000\\
Sun to Moon  & 100,000 \\
Moon to Asterisms  & 100,000 \\
Asterisms to Mercury  & 200,000 \\
Mercury to Venus  & 200,000 \\
Venus to Mars  & 200,000 \\
Mars to Jupiter  & 200,000 \\
Jupiter to Saturn  & 200,000 \\
Saturn to Ursa Major  & 100,000 \\
Ursa Major to Pole-star  & 100,000 \\\hline
Sub-total  & 17,100,000 \\ \hline
\end{tabular}
\vspace{0.3in}

Further spheres are postulated 
beyond the pole-star.
These are the Maharloka, the Janaloka, the Tapoloka, and the
Satyaloka. Their distances are as follows:

\vspace{0.3in}
Table 2: From Pole-star to Satyaloka

\begin{tabular}{||l|r||} \hline
Interval II & yojanas\\ \hline
Pole-star to Maharloka  & 10,000,000 \\
Maharloka to Janaloka  & 20,000,000 \\
Janaloka to Tapoloka  & 40,000,000 \\
Tapoloka to Satyaloka  & 120,000,000 \\\hline
Grand Total  & 207,100,000 \\ \hline
\end{tabular}

\vspace{0.3in}

Since the last figure is the distance from the Earth, the
total diameter of the universe is 414.2 million yojanas,
not including the dimensions of the various heavenly bodies
and {\it lokas}.
The inclusion of these may be expected to bring this
calculation in line with the figure of 500 million
yojanas mentioned earlier.

Beyond the universe lies the limitless {\it Pradh\={a}na},
that has within it countless other universes.

Pur\={a}\d{n}ic cosmology views the universe as going
through cycles of creation and destruction of 8.64 billion years.
The consideration of a universe of enormous size must have
been inspired by a supposition of enormous age.

\subsection*{Reconciling Pur\={a}\d{n}ic and Standard Yojanas}

It is clear that the Pur\={a}\d{n}ic yojana ($y_p$) are
different from the Artha\'{s}\={a}stra yojana ($y_p$).
To find the conversion factor, we equate the distances to
the Sun.

\begin{equation}
0.4375 \times 10^6 ~y_s = 15.7 \times 10^6 ~y_p
\end{equation}

In other words,

\begin{equation}
1 ~y_s \approx 36 ~y_p
\end{equation}

The diameter of the Earth should now be about
$875 \times 36 \approx 31,500 ~ y_p$.
Perhaps, this was taken to be 32,000 $y_p$,
twice the size of Meru.
This understanding is confirmed by the statements in
the Pur\={a}\d{n}as.
For example, MP 126 says that 
the size of Bh\={a}ratavar\d{s}a (India)
is 9,000 $y_p$, 
which is roughly correct.

We conclude that the kernel of the Pur\={a}\d{n}ic system
is consistent with the Siddh\={a}ntas.
The misunderstanding of it arose because attention was
not paid to their different units of distance.

\section*{Speed of the Sun}

Now that we have a Pur\={a}\d{n}ic context,
S\={a}ya\d{n}a's statement on the speed
of 2,202 
yojanas per half-nime\d{s}a can be examined.

We cannot be absolutely certain what yojanas did he have in mind:
standard, or Pur\={a}\d{n}ic.
But either way it is clear from the summary of Pur\={a}\d{n}ic
cosmology that this speed could not be the speed of the Sun.
At the distance of 15.7 million yojanas, Sun's speed 
is only 121.78 yojanas ($y_p$) per half-nime\d{s}a.
Or if we use the the figure from VaP,
it is 116.67.
Converted into the standard yojanas, this number
is only 3.24 $y_s$ per half-nime\d{s}a.

S\={a}ya\d{n}a's speed is about 18 times
greater than the supposed speed of the Sun in $y_p$
and $2 \times 18^2$ greater than the speed in $y_s$.
So either way, a larger number with a definite
relationship to the actual speed of the Sun was chosen
for the speed of light.

The Pur\={a}\d{n}ic size of the universe is 13 to 16
times greater than the orbit of the Sun, not counting
the actual sizes of the various heavenly bodies.
Perhaps, the size was taken to be 18 times greater
than the Sun's orbit.
It seems reasonable to assume, then, that if the
radius of the universe was taken to be about 282 million
yojanas, a speed was postulated for light so that
it could circle the farthest path in the universe within
one day.
This was the physical principle at the basis of
the Pur\={a}\d{n}ic cosmology.

\section*{Concluding Remarks}

We have seen that the astronomical 
numbers in the Pur\={a}\d{n}as are much more consistent 
amongst themselves, and with
the generally accepted 
sizes of the solar orbit, than has been hitherto assumed.
The Pur\={a}\d{n}ic geography must not be taken
literally.

We have also shown that
the S\={a}ya\d{n}a's figure of 2,202 yojanas
per half-nime\d{s}a is consistent with
Pur\={a}\d{n}ic cosmology where the
size of ``our universe'' is a function of
the speed of light.
This size represents the space that can be
spanned by light in one day.

It is quite certain that the figure for speed was obtained either by
this argument or it was obtained by taking the
postulated speed of the Sun in the Pur\={a}\d{n}as and
multiplying that by 18, or by multiplying the
speed in standard yojanas by $2 \times 18^2$.
We do know that 18 is a sacred number in the Pur\={a}\d{n}as,
and the fact that multiplication with this special number gave
a figure that was in accord with the spanning of light in the
universe in one day must have given it a special significance.

Is it possible that the number 2,202 arose because of a mistake
of multiplication by 18 rather than a corresponding division
(by 36) to reduce the Sun speed to standard yojanas? The answer
to that must be ``no'' because such a mistake is too egregious.
Furthermore, S\={a}ya\d{n}a's own brother M\={a}dhava was a
distinguished astronomer and the incorrectness of this figure
for the accepted speed of the Sun would have been obvious to him.

If S\={a}ya\d{n}a's figure was derived from a postulated size of the 
universe, how was that huge size, so central
to all Indian thought, arrived at?
A possible explanation is that the
physical size of the universe was taken to parallel
the estimates of its age.
These age-estimates were made larger and larger to 
postulate a time when the periods of all the
heavenly bodies were synchronized.$^{12}$

The great numbers of the Pur\={a}\d{n}as suggest
that the concepts of mah\={a}yuga and
kalpa, sometimes credit to the astronomers of the
Siddh\={a}ntic period, must have had an old
pedigree.
This is in consonance with the new understanding
that considerable astronomy was in place in the
second and the third millennia BC.$^{13}$

We have provided a context in which S\={a}ya\d{n}a's speed
can be understood.
In this understanding, the speed of light was
taken to be $2 \times 18^2$ greater than the speed of
the Sun in standard yojanas so that
light can travel the entire postulated size of the universe
in one day. It is a lucky chance that
the final number turned out to be exactly equal to
the true speed.
S\={a}ya\d{n}a's value as speed of light must be considered
the most astonishing ``blind hit'' in the history of science!

\section*{Notes}
\begin{enumerate}

\item M\"{u}ller, Max (ed.), {\it Rig-Veda-Samhita together
with the Commentary of S\={a}ya\d{n}a.} Oxford University Press, London, 1890.

\item De, S.S., In {\it Issues in Vedic Astronomy and Astrology}, Pandya, H,
Dikshit, S., Kansara, M.N. (eds.). Motilal Banarsidass, Delhi, 1992, pages
234-5;\\
Vartak, P.V., {\it Scientific Knowledge in the Vedas.}
Nag Publishers, Delhi, 1995.\\
See also, Kak, S.C., 1998. 
{\em Indian Journal of History of Science,} 33, 31-36.

\item Shrava, S., {\it History of Vedic Literature.} Pranava Prakashan, New
Delhi, 1977, p. 185.

\item Larson, G.J. and
Bhattacharya, R.S. (ed.), {\it S\={a}\d{m}khya: A Dualist Tradition in
Indian Philosophy,}
Princeton University Press, Princeton, 1987;\\
Matilal, B.K., {\it Ny\={a}ya-Vai\'{s}e\d{s}ika,}
Otto Harrassowitz, Wiesbaden, 1977;\\
Potter, K.H. (ed.), {\it Indian Metaphysics and Epistemology,}
Princeton University Press, Princeton, 1977.
 
\item Seal, B., {\it The Positive Sciences of the Hindus.}
Motilal Banarsidass, Delhi, 1985 (1915)

\item Kak, S.C., 1998. 
{\em Indian Journal of History of Science,} 33, 93-100.

\item Shukla, K.S. and Sarma, K.V., {\it
\={A}ryabha\d{t}\={\i}ya of \={A}ryabha\d{t}a.}
Indian National Science Academy, New Delhi, 1976.

\item Shukla, K.S., {\it
\={A}ryabha\d{t}\={\i}ya of \={A}ryabha\d{t}a with the
Commentary of Bh\={a}skara I and Some\'{s}vara.}
Indian National Science Academy, New Delhi, 1976, pp. 26-27.

\item  Kak, S.C., {\em The Astronomical Code of the \d{R}gveda}. 
Aditya, New Delhi, 1994.

\item
Rocher, L., {\it The Pur\={a}\d{n}as}.
Otto Harrassowitz, Wiesbaden, 1986;\\
Wilson, H.H. (tr.), {\it The Vishnu Purana}.
Trubner \& Co, London, 1865 (Garland Publishing, New York, 1981);\\
{\it The Matsya Puranam}. The Panini Office, Prayag, 1916 (AMS, New
York, 1974);\\
Tripathi, R.P. (tr.),  {\it The V\={a}yu Pur\={a}\d{n}a}. Hindi
Sahitya Sammelan, Prayag, 1987.

\item de Santillana, G. and von Dechend, H., {\it Hamlet's Mill: An Essay
on Myth and the Frame of Time.} Gambit, Boston, 1969.

\item Kak, S.C.,
{\em Vistas in Astronomy,} 36, 117-140, 1993.

\item  Kak, S.C., 
{\em Quarterly Journal of the Royal Astronomical Society,} 36, 385-396, 1995;
Kak, S.C., 
{\em Quarterly Journal of the Royal Astronomical Society,} 37, 709-715, 1996.

\comment{
\item Neugebauer, O., {\it A History of Ancient Mathematical
Astronomy}. Springer-Verlag, Berlin, 1975.

\item  Kak, S.C., 
"The astronomical of the age of geometric altars", 
{\em Quarterly Journal of the Royal Astronomical Society,} 36, 385-396, 1995.

\item  Kak, S.C., 
"Knowledge of the planets in the third millennium BC", 
{\em Quarterly Journal of the Royal Astronomical Society,} 37, 709-715, 1996.

\item 
Kak, S.C., ``The sun's orbit in the Br\={a}hma\d{n}as,''
{\em Indian Journal of History of Science,} in press;

Kak, S.C., ``The astronomy of the \'{S}atapatha Br\={a}hma\d{n}a,''
{\em Indian Journal of History of Science,} 28, 15-34, 1993.

\item  Kak, S.C., {\em The Astronomical Code of the \d{R}gveda}. 
Aditya, New Delhi, 1994.

\item Thurston, H., {\it Early Astronomy}.
Springer-Verlag, New York, 1994, page 188.

\item Neugebauer, 1975, page 7.

\item Burgess, E., {\it The S\={u}rya Siddh\={a}nta.}
Motilal Banarsidass, Delhi, 1989 (1860), pages 389-390.
}

\end{enumerate}

\end{document}